\documentclass[conference, 12pt]{IEEEtran}
\usepackage[mathcal]{euscript}
\usepackage{amsmath}
\usepackage{amsthm}
\usepackage{amssymb}
\usepackage{graphicx}
\usepackage{algorithmic}
\usepackage{algorithm}
\usepackage{epsfig}
\usepackage{booktabs}
\usepackage{colortbl}
\usepackage{amsmath}
\usepackage{xcolor}
\usepackage{graphicx}
\usepackage{cite} 

\providecommand{\mbf}[1]{\mathbf{#1}}						 
\providecommand{\bsym}[1]{\boldsymbol{#1}}				 

\providecommand{\mbb}[1]{\mathbb{#1}}	

\begin{document}

\title{On The Impact of Time-Varying Interference-Channel on the Spatial Approach of Spectrum Sharing between S-band Radar and Communication System}

\author{\begin{tabular}{c}
 Awais Khawar, Ahmed Abdelhadi, and T. Charles Clancy \\
\{awais, aabdelhadi, tcc\}@vt.edu \\
Ted and Karyn Hume Center for National Security and Technology\\
Bradley Department of Electrical and Computer Engineering\\
Virginia Tech, Arlington, VA, 22203, USA
\end{tabular}}

\maketitle

\begin{abstract}
Spectrum sharing is a new approach to solve the congestion problem in RF spectrum.  
A spatial approach for spectrum sharing between radar and communication system was proposed, which mitigates the radar interference to communication by projecting the radar waveform onto null space of interference channel, between radar and communication system \cite{SKC+12}. In this work, we extend this approach to maritime MIMO radar which experiences time varying interference channel due to the oscillatory motion of ship, because of the breaking of sea/ocean waves. We model this variation
by using the matrix perturbation theory and the statistical distribution of the breaking waves. This model is then used to study the impact of perturbed interference channel on the spatial approach of spectrum sharing. We use the maximum likelihood (ML) estimate of target's angle of arrival to study the radar's performance when its waveform is projected onto the null space of the perturbed interference channel. Through our analytical and simulation results, we study the loss in the radar's 
performance due to the null space projection (NSP) of its waveform on the perturbed interference channel.


\end{abstract}


\section{Introduction}
Spectrum sharing between radars and communication system is a new way forward to solve the growing problem of spectrum congestion \cite{PCAST12}. The 3550-3650 MHz band is identified for spectrum sharing between military radars and broadband wireless access (BWA) communication systems, in the National Telecommunications and Information Administration's (NTIA) 2010 Fast Track Report \cite{NTIA10}. 

Electromagnetic interference (EMI) to radars and BWA systems is expected from spectrum sharing \cite{KAC14WTS}. In this paper, we focus on mitigating radar's interference to communication systems. The emission pattern of radar, especially high transmit power and high-peak sidelobes, saturates communication system receiver, which traditionally operate at very small power levels. However, radar's interference can be mitigated at communication systems by exploiting the advancements in transmitter and receiver designs, and through a combination of spatial and temporal signal processing algorithms.


Spatial algorithms for interference mitigation is a well explored topic in the cognitive radio research community \cite{Yi10}, however, a spatial algorithm to mitigate radar's interference to communication system was first proposed in \cite{SKC+12}. The waveform of radar is projected onto null space of interference channel between the radar and the communication system. This NSP of the radar waveform mitigates radar interference but in turn causes slight degradation in the radar performance, which is studied in \cite{SKC+12}. 

For a maritime radar, i.e., a radar mounted on a ship, interference channel between the radar and communication system is subject to change due to many factors, motion of the ship due to waves is one of them. In this paper, we focus on the accuracy of available interference-channel-state-information (ICSI) for the NSP spectrum-sharing algorithm. We model the imperfection or uncertainty in the ICSI by using the matrix perturbation theory. The NSP algorithm is sensitive to perturbations in the ICSI as it can alter the null space of the interference channel. This can cause interference leakage to a communication system, if it is not accounted for, and it can also degrade the performance of a radar system, which is addressed in this paper. Our goal is to study the impact of perturbed ICSI on the NSP algorithm and the radar performance. 

\textit{\textbf{Relation to prior work:}}
Prior work on the NSP technique to share RF spectrum between radar and communication system focuses on the assumption that perfect ICSI is available at the radar \cite{SKC+12}. However, this assumption is impractical since ICSI can have variations due to errors in the estimation process, quantization of feedback, or time variations in interference-channel, due to motion. In our case, we consider variation in interference-channel due to the up-and-down motion of the ship, because of waves, and not due to the communication system, which we assume is fixed. Thus, this brings in imperfect ICSI at the radar which is not considered in \cite{SKC+12}. We study the impact of an inaccurate ICSI
on the performance of radar.

\textit{\textbf{Notations:}} Bold capital letters, e.g., $\mbf A$, correspond to matrices and bold letters, e.g., $\mbf a$, correspond to column vectors. The operators $(\cdot)^H$ and $(\cdot)^*$ corresponds to the Hermitian transpose and complex-conjugate, respectively.

The remainder of this paper is organized as
follows. Section~\ref{sec:arch} discusses spectrum sharing architecture.
Section \ref{sec:radar} briefly discusses the theory of colocated MIMO radar. Section \ref{pertmodel} models perturbed ICSI.
In Section \ref{sec:NSP}, we study perturbed channel's impact on the NSP algorithm and the ML estimate. 
Section~\ref{sec:sim}
discusses the simulation setup and provides quantitative results along with the discussion. Section~\ref{sec:conc} concludes the paper. 

\begin{figure}
\centering
\includegraphics[trim=0in 0.8in 0in 0in,width=3.2in]{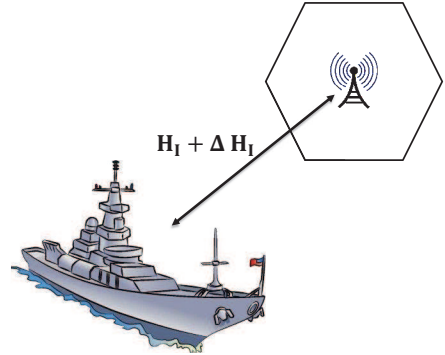}
\caption{A maritime MIMO radar sharing spectrum with a MIMO communication system. The interference channel $\mbf H_{\text{I}}$ is perturbed by $\Delta \mbf H_{\text{I}}$ due to the motion of the ship.} 
\label{fig:icassp}
\end{figure}

\section{Spectrum Sharing Architecture}\label{sec:arch}

In this paper, we consider a MIMO communication system equipped with $N_T$ transmit antennas and $N_R$ receive antennas sharing the 3550-3600 MHz RF band with a MIMO radar with $M_T$ transmit antennas and $M_R$ receive antennas. 

We assume ICSI of the communication system, or its distribution, is available at the radar system. In practice, this assumption can be justified when both the radar and communication systems belong to military. In this case, let the received signal at the communication system's receiver be
\begin{equation*}
\mbf y(t) = \mbf H_{\text{I}}^{N_R \times M_T} \mbf x_{\text{Radar}}(t) + \mbf H^{N_R \times N_T} \mbf x(t) + \mbf n(t)
\end{equation*}
where $\mbf H_{\text{I}}$ is the ${N_R \times M_T}$ interference channel between the radar and the communication system \cite{KMA_2DChannelModel}, $\mbf H$ is the ${N_R \times N_T}$ communication channel between the communication systems, and $\mbf n(t)$ is the additive white Gaussian noise. In order to avoid the interference to the communication system, the spatial approach is to project radar signal $\mbf x_{\text{Radar}}(t)$ onto the null-space of $\mbf H_{\text{I}}$ such that $\mbf H_{\text{I}} \mbf x_{\text{Radar}}(t) = \bsym 0$. 

However, due to oscillatory motion of the ship, because of sea/ocean waves even if the ship is docked at harbor, the interference channel between the radar and the communication system is perturbed. In this case, the received signal at the communication system's receiver is
\begin{equation*}
\mbf y(t) = [\mbf H_{\text{I}} + \Delta \mbf H_{\text{I}}] \mbf x_{\text{Radar}}(t) + \mbf H \mbf x(t) + \mbf n(t).
\end{equation*}
where $\Delta \mbf H_{\text{I}}$ denotes perturbation in ICSI and is modeled using Rayleigh distribution, see Section \ref{pertmodel} for details, with covariance matrix
\begin{equation*}
\text{Cov}_{\Delta \mbf H_{\text{I}}} = \mbb E \left\{ vec(\Delta \mbf H_{\text{I}}) vec(\Delta \mbf H_{\text{I}})^H \right\}
\end{equation*}
where $vec(\cdot)$ is the column stacking operator. This spectrum sharing scenario is shown in Figure \ref{fig:icassp}.

As previously stated, spatial approach of interference mitigation is projection of radar signal onto the null space of interference channel \cite{KAC+14DySPANProjection}. After perturbation, in order for the spatial approach to work, the radar signal is projected onto the null space of the perturbed interference channel such that $[\mbf H_{\text{I}} + \Delta \mbf H_{\text{I}}] \mbf x_{\text{Radar}}(t) = \bsym 0$. 




\section{Colocated MIMO Radar}\label{sec:radar}
In this section, we introduce preliminaries of colocated MIMO radar, having $M_T$ transmit and $M_R$ receive antennas, which transmits finite-alphabet constant-envelop BPSK waveforms designed in \cite{KAC+14ICNC}. The radar waveform $\mbf x_{\text{Radar}}(t)$ can be expressed as 
\begin{equation*}
\mbf x_{\text{Radar}}(t)= \begin{bmatrix} x_1(t)e^{j \omega_c t} &x_2(t)e^{j \omega_c t} &\cdots &x_{M_T}e^{j \omega_c t}(t) \end{bmatrix}^T
\end{equation*} 
where $x_k(t)$ is the baseband signal from the $k^{\text{th}}$ transmit element, $\omega_c$ is the carrier angular frequency,
%
$t \in [0, T_o]$, with $T_o$ being the observation time. Then the received signal, from a single point target at an angle $\theta$, is 
\begin{equation*}
\mbf y_{\text{Radar}}(t) = \alpha \, \mbf A(\theta) \,  \mbf x_{\text{Radar}}(t-\tau(t)) + \mbf n(t)
\end{equation*}
where $\mbf n(t)$ is additive white Gaussian noise, $\tau(t)=\tau_{T_k}(t) + \tau_{R_l}(t)$, denoting the sum of propagation delays between the target and the $k^{\text{th}}$ transmit element and the $l^{\text{th}}$ receive element, respectively; and $\alpha$ represents the complex path loss including the propagation loss and the reflection coefficient; $\mbf a_T(\theta)$ is the transmit steering matrix defined as
\begin{equation*}
\mbf a_T(\theta) \triangleq \begin{bmatrix} e^{-j \omega_c \tau_{T_1}(\theta)} &e^{-j \omega_c \tau_{T_2}(\theta)} &\cdots &e^{-j \omega_c \tau_{T_{M_T}}(\theta)} \end{bmatrix}^T
\label{eq:at}
\end{equation*}
$\mbf a_R(\theta)$ is the receive steering matrix defined as
\begin{equation*}
\mbf a_R(\theta) \triangleq \begin{bmatrix} e^{-j \omega_c \tau_{R_1}(\theta)} &e^{-j \omega_c \tau_{R_2}(\theta)} &\cdots &e^{-j \omega_c \tau_{R_{M_R}}(\theta)} \end{bmatrix}^T
\label{eq:ar}
\end{equation*}
and $\mbf A (\theta)$ is the transmit-receive steering matrix defined as
\begin{equation*}
\mbf A (\theta) \triangleq \mbf a_R(\theta) \mbf a_T^T(\theta).
\end{equation*}

Our performance metric for the MIMO radar's performance is the maximum likelihood (ML) estimate of the target's angle of arrival, which is expressed as, as in \cite{LS08},
\begin{equation}
\label{eq:ML}
(\hat{\theta},\hat{\tau}_r, \hat{\omega}_D)_{\text{ML}} = \operatorname*{arg\,max}_{\theta,\tau_r,\omega_D} \frac{\left| \mathbf{a}_R^H(\theta) \mbf E(\tau_r,\omega_D)\mathbf{a}_T^*(\theta)\right|^2}{M_R \mathbf{a}_T^H(\theta) \mathbf{R}_{\mbf x_{\text{Radar}}}^T \mathbf{a}_T(\theta)}
\end{equation}
where 
\begin{align*}
\mbf R_{\mbf x_{\text{Radar}}} &= \int_{T_0} \mbf x_{\text{Radar}}(t) \, \mbf x_{\text{Radar}}^H(t) \, dt \\
\mbf E(\tau_r,\omega_D) &= \int_{T_0} \! \mathbf{y}_{\text{Radar}}(t) \, \mbf x_{\text{Radar}}^H(t-\tau_r) \, e^{j\omega_D t} \, dt
\end{align*}
where $\tau_r$ is the propagation delay between the target and the reference point, and $\omega_D$ is the Doppler frequency shift as defined in Table 2 \cite{SKC+12}.

\section{Physical Modeling Of The Perturbed Interference Channel} \label{pertmodel}

In this paper, the cause of perturbation in ICSI is due to the motion of the maritime radar, mounted on ship. This motion is induced by sea/ocean waves. So, in order to statistically describe the perturbation we look at the statistics of the wave height. The probability density function (pdf) of the wave height $h$ is given by the Rayleigh distribution as, as in \cite{LH52},
\begin{equation*}
p(h) = \frac{2h}{h_{\text{rms}}^2} \exp \left[ - \left( \frac{h}{h_{\text{rms}}} \right)^2 \right]
\end{equation*}
where $h_{\text{rms}}$ is the root-mean-square (rms) wave height defined as
\begin{equation*}
h_{\text{rms}} = \left[ \frac{1}{N} \sum_{n=1}^N h_n^2 \right]^{1/2}
\end{equation*}
where $N$ corresponds to the number of observed waves and the $n^{\text{th}}$ wave has height $h_n$. Thus, $h_{\text{rms}}$ can be computed by observing $N$ wave heights. We choose Rayleigh pdf for the wave height distribution as it gives a very good estimate of the wave properties for various nearshore conditions.

\section{Impact of The Perturbed Interference Channel on the NSP Algorithm}\label{sec:NSP}
\label{sec:nsp}
The projection of the radar signals onto the null space of the interference channel via a projection algorithm was proposed in \cite{SKC+12} for a stationary interference channel. However, due to the motion of the ship, the interference channel gets perturbed and can be written as
\begin{equation*}
{\mbf H}_{ \Delta \text{I}} = \mbf H_{\text{I}} + \Delta \mbf H_{\text{I}}.
\end{equation*}
For the NSP algorithm, the null space of the perturbed interference channel can be calculated from the singular value decomposition (SVD) theorem as
\begin{equation*}
{\mbf H}_{ \Delta \text{I}} = \mathbf{U}_{ \Delta \text{I}} \mathbf{\Sigma}_{ \Delta \text{I}} \mathbf{V}_{ \Delta \text{I}}^{H}
\end{equation*}
where $\mathbf{U}_{ \Delta \text{I}}$ is the perturbed complex unitary matrix, $\mathbf{\Sigma}_{ \Delta \text{I}}$ is the diagonal matrix of the perturbed singular values, and $\mathbf{V}_{ \Delta \text{I}}^{H}$ is the perturbed complex unitary matrix and its columns corresponding to the vanishing singular values span the null space of ${\mbf H}_{ \Delta \text{I}}$. This is denoted by $\breve{\mathbf{V}}_{ \Delta \text{I}}$. We project the radar signal onto the null space of the perturbed interference-channel using the projection algorithm, as in \cite{SKC+12},
\begin{eqnarray*}
\mathbf{P}_{\breve{\mathbf{V}}_{ \Delta \text{I}}}=\breve{\mathbf{V}}_{ \Delta \text{I}}\breve{\mathbf{V}}_{ \Delta \text{I}}^H. \label{Pv}
\end{eqnarray*}
The radar waveform projected onto the null space of ${\mbf H}_{ \Delta \text{I}}$ can be written as
\begin{equation}
\label{eq:news}
{\breve {\mbf x}_{\text{Radar}}} = \mathbf{P}_{\breve{\mathbf{V}}_{ \Delta \text{I}}} \mbf x_{\text{Radar}}.
\end{equation}
By inserting the projected signal, equation \eqref{eq:news} in equation \eqref{eq:ML}, we get the ML estimate of the target's angle of arrival for the NSP projected radar waveform as
\begin{equation}
\label{eq:MLnsp}
(\hat{\theta},\hat{\tau}_r, \hat{\omega}_D)_{\text{ML}_{\text{NSP}}} = \operatorname*{arg\,max}_{\theta,\tau_r,\omega_D} \frac{\left| \mathbf{a}_R^H(\theta) \mbf E(\tau_r,\omega_D)\mathbf{a}_T^*(\theta)\right|^2}{M_R \mathbf{a}_T^H(\theta) \mathbf{R}_{\breve {\mbf x}_{\text{Radar}}}^T \mathbf{a}_T(\theta)} \cdot
\end{equation}


\section{Simulation Results}\label{sec:sim}
In this section, we simulate the impact of time varying interference channel on MIMO radar-communication system spectrum-sharing scenario. The interference channel is modeled to have a Rayleigh distribution. The elements of the error channel $\Delta \mbf H_{\text{I}}$ are modeled as Rayleigh distribution, as it describes the pdf of the height of the waves, with $h_{\text{rms}}$ taking values 1, 2, 3, and 4, indicating perturbation 
in the channel coefficients. The ML results are based on the average of 10,000 independent trials. The simulation parameters used are listed in Table 2 \cite{SKC+12}.



\begin{figure}
\centering
	\includegraphics[width=\linewidth]{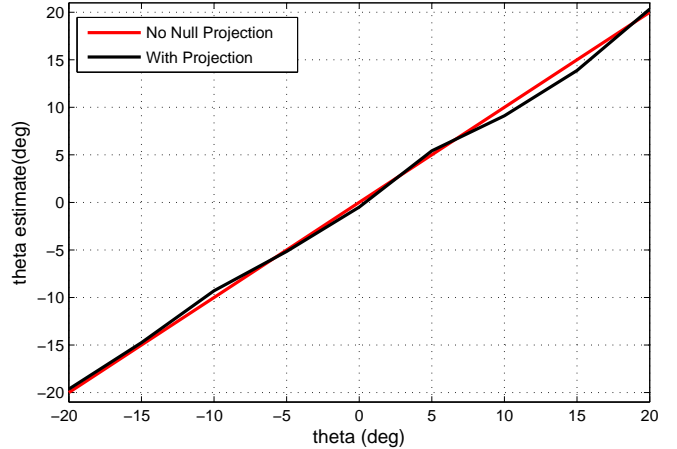} 
	\caption{The ML estimate of target's angle of arrival. The performance of the original waveform is compared with the null space projected waveform.}
		\label{fig:ml}
\end{figure}

\begin{figure*}
\centering
	\includegraphics[width=\linewidth]{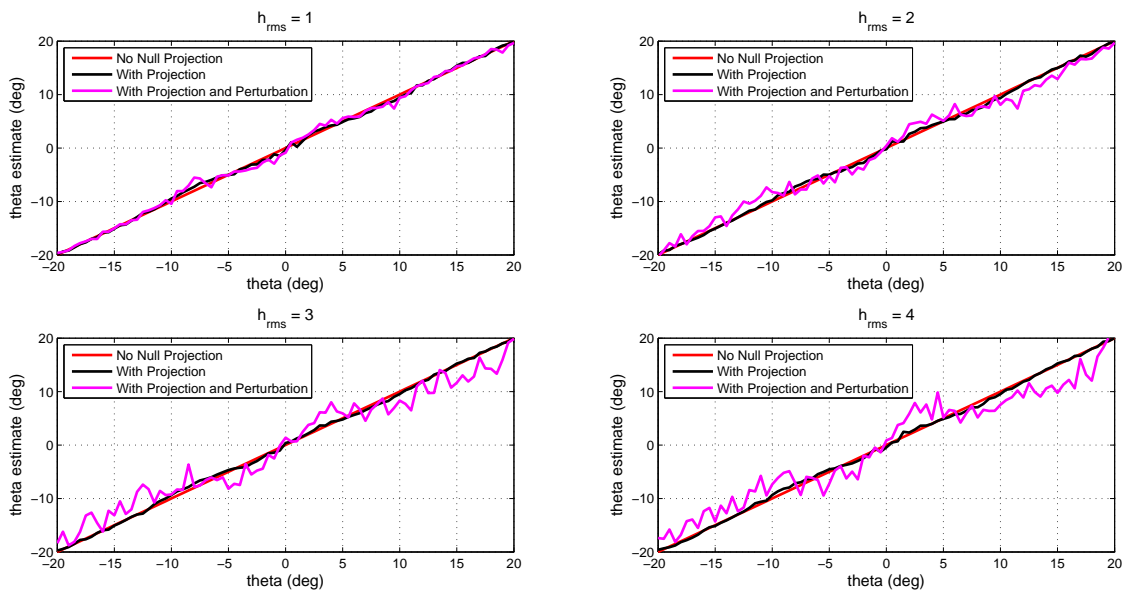} 
	\caption{The ML estimate of target's angle of arrival. The performance of the original waveform is compared with the null space projected waveform onto the perturbed and un-perturbed interference channel. The values of the rms wave height, i.e., $h_{\text{rms}} =1$, $h_{\text{rms}} =2$, $h_{\text{rms}} =3$, and $h_{\text{rms}} =4$ indicate a perturbation in the channel coefficients.}
		\label{fig:pert}
\end{figure*}

The ML estimate of the target's angle of arrival is calculated by equations \eqref{eq:ML} and \eqref{eq:MLnsp} for the original radar waveform and the NSP radar waveform on the perturbed interference-channel, respectively. The impact of the NSP of the radar waveform, on the perturbed interference-channel, can be studied by the ML estimation error of the angle of arrival. First, in Figure \ref{fig:ml}, we compare the original and the estimated angles using the ML estimation for the original radar waveform and the NSP waveform onto the un-perturbed interference channel. It is important to note that the NSP waveform doesn't degrade the ML estimates or radar performance. This endorses the claims of \cite{SKC+12} that the NSP for spectrum-sharing is a viable approach for MIMO radars. Second, we analyze the effect of the perturbation on the NSP algorithm. In Figure \ref{fig:pert}, we compare the original and the estimated angles for various magnitudes of the perturbation in the channel coefficients. It is important to note that due to the perturbation of the interference-channel the degradation in the performance of the NSP waveform is significant. The degradation also depends on the magnitude of the perturbation. It is evident that as the perturbation in the channel coefficients increases the performance of the ML, for the NSP waveform, degrades significantly. Thus, perturbation in the interference channel can have a detrimental effect on the radar performance, in a spectrum-sharing setting, when NSP approaches are used. This can be avoided by having a frequent exchange of the ICSI between radars and communication systems. 



\section{Conclusion}\label{sec:conc}

In future, federal radar spectrum will be shared with commercial operators. In this paper, we explore the spectrum-sharing scenario when the interference channel between a radar and a communication system is subject to variations due to the oscillatory motion of ship. 
We study the effect of perturbation, in the interference channel, when the NSP approach of spectrum sharing is used. The NSP algorithm does not degrade the radar performance when complete ICSI is available, however, when the interference channel is perturbed, by the waves, the radar's performance is degraded. The magnitude of the degradation depends upon the magnitude of the perturbation in the channel coefficients, which is due to the height of the waves. The radar performance, in such a scenario, can be improved by having frequent updates of the ICSI.

\section{Future Work}\label{sec:future}

In this paper, we considered only a single communication system sharing spectrum with a radar. However, we are looking into extending this work to multi-cell communication system \cite{KAC+14DySPANProjection} and studying target detection performance of spectrum sharing MIMO radars \cite{KAC14_TDetect}. Moreover, instead of using orthogonal waveforms, MIMO radar waveforms with spectrum sharing constraints can be designed which can mitigate radar interference to communication systems \cite{KAC14DySPANWaveform, KAC14_QPSK}. 
%
%
%
%
%
%
%
%
%
%
%
%
We are also working on various algorithms that exploit the NSP, at radar, and resource allocation along the lines of \cite{Ahmed_Utility1, Ahmed_Utility4, AKC14_PowerAllocation, Haya_Utility1, Haya_Utility2, SKA+14DySPAN, MAC14}, at communication system, to mitigate interference in a radar-communication system spectrum-sharing scenario.



\bibliographystyle{ieeetr}
\bibliography{IEEEabrv,MILCOMpaper}

\end{document}